\documentclass[pdftex,twocolumn,epjc3]{svjour3}          

\RequirePackage[T1]{fontenc}

\smartqed  

\RequirePackage{graphicx}
\RequirePackage{mathptmx}      
\RequirePackage{flushend}
\RequirePackage[numbers,sort&compress]{natbib}
\RequirePackage[colorlinks,citecolor=blue,urlcolor=blue,linkcolor=blue]{hyperref}

\journalname{Eur. Phys. J. C}

\begin{document}

\title{Explaining low $\ell$ Anomalies in the CMB Power Spectrum with Resonant Superstring Excitations during 
Inflation}


\author{Mayukh~R.~Gangopadhyay\thanksref{e1,addr1,addr2}
        \and
        Grant~J.~Mathews\thanksref{e2,addr1,addr3} 
        \and
        Kiyotomo Ichiki\thanksref{e3,addr4}
        \and
        Toshitaka Kajino\thanksref{e4,addr3,addr5,addr6}
}

\thankstext{e1}{e-mail: mayukh.raj@saha.ac.in }
\thankstext{e2}{e-mail: gmathews@nd.edu}
\thankstext{e3}{e-mail: ichiki.kiyotomo@c.mbox.nagoya-u.ac.jp}
\thankstext{e4}{e-mail: kajino@nao.ac.jp}

\institute{Center for Astrophysics, Department of Physics,University of Notre Dame, Notre Dame, IN 46556, USA \label{addr1}
          \and
          Theory Division, Saha Institute of Nuclear Physics, 1/AF, Bidhannagar, Kolkata- 64, India\label{addr2}
          \and
         National Astronomical Observatory Of Japan, 2-21-1, Osawa, Mitaka, Tokyo 181-8588, Japan \label{addr3}
         \and
         Department of Physics, Nagoya University, Nagoya 464-8602, Japan\label{addr4}
         \and
       Graduate School of Science, The University of Tokyo, 7-3-1 Hongo, Bunkyo-ku, Tokyo 113-0033, Japan\label{addr5}
         \and
         International Research Center for Big-Bang Cosmology and Element Genesis, and School of Physics and Nuclear Energy Engineering, Beihang University, Beijing 100083, China\label{addr6}}

\date{Received: date / Accepted: date}

\maketitle

\begin{abstract}
We explore the possibility that both the suppression of the $\ell = 2$ multipole moment of the power spectrum of cosmic microwave background temperature fluctuations and the possible dip for $\ell = 10-30$ can be explained as well as a possible new dip for $\ell \approx 60$ as the result of the resonant creation of  sequential excitations of a fermionic (or bosonic) closed superstring that couples to the inflaton field. We consider a D=26 closed bosonic string with one toroidal compact dimension as an illustration of how string excitations might imprint themselves on the CMB. We analyze the existence of successive momentum states, winding states or oscillations on the string as the source of the three possible dips in the power spectrum. Although the evidence of these dips are of marginal statistical significance, this might constitute the first observational evidence of successive superstring excitations in Nature.
 \end{abstract}

\section{Introduction}
\label{sec:intro}

It is generally accepted that the energy scale of superstrings is so high that it is impossible to ever observe a superstring in the laboratory.  There is, however, one epoch in which the energy scale of superstrings was obtainable in Nature.  That is in the realm of the early moments of trans-Plankian \cite{Martin01} chaotic inflation out of the string theory landscape.  

There have been a number of papers exploring the possible impact of string theory on the cosmic microwave background \cite{vafaei}-\cite{Mathews15}. This paper explores the possibility that a specific sequence of super-string excitations may have made itself known via its coupling to the inflaton field of inflation.  This may have left an imprint of "dips" \cite{Mathews15} in the $TT$ and $EE$ power spectra of the cosmic microwave background.  The identification of this particle as a superstring is possible because there may be evidence for  sequential oscillator states of the same superstring that appear on different scales of the sky.  Nevertheless, the point of this paper cannot possibly be to provide the final formulation of a string theoretic explanation for deviations in the  CMB power at low multipoles within a model that is fully realistic as a particle physics model. The aim of this paper is rather to point out a potentially interesting cosmology that may have an implication in a deeper string theory.  Our goal is to provide a proof-of-principle within a model that has most of the relevant coarse features or a realistic string theory in hopes that this could inspire further investigation.

The  primordial power spectrum is believed to derive from quantum fluctuations 
generated
during the inflationary epoch \cite{Liddle,cmbinflate}.  The various observed power spectra of the cosmic microwave background (CMB) are  then modified by the dynamics of the cosmic radiation and matter  fluids as various scales  re-enter the horizon along with effects from the transport of photons from the epoch of last scattering to the present time.  
Indeed,   the {\it Planck} data \cite{PlanckXIII,PlanckXX} have provided the highest resolution yet available in the  determination of
CMB power spectra.  Although the {\it TT} primordial power spectrum is well fit with a simple tilted power law \cite{PlanckXX}, there remain at least two interesting features that may suggest deviations from the simplest inflation  paradigm.  

One such feature is the  well known suppression of the $\ell=2$ moment of the CMB power spectrum observed both by {\it Planck} \cite{PlanckXIII} and by the Wilkinson Microwave Anisotropy Probe ({\it WMAP}) \cite{WMAP9}.  There is also a feature of marginal statistical significance \cite{PlanckXX} 
in the observed power spectrum of both {\it Planck} and {\it WMAP} 
near multipoles $\ell = 10-30$.  Both of these deviations occur in an interesting region in the CMB power spectrum because they correspond to 
angular scales that were not yet in causal contact when the CMB photons were emitted.  Hence, the observed power spectra  close to  the true primordial power spectrum for these features.  

In the {\it Planck} 
inflation parameters paper \cite{PlanckXX}, however,   
the deviation from a simple power law  in the range $\ell = 10-30$ was deduced to be of  weak  statistical significance due to the large cosmic variance at low $\ell$.  
In particular, a range of models was considered from the minimal case of a kinetic energy dominated phase preceding a short inflationary stage (with just one extra parameter), to a model with a step-like feature in the inflation generating potential and in the sound speed (with five extra parameters). These modifications led to  improved fits of up to $\Delta \chi^2 = 12$. However,  neither the Bayesian evidence nor a frequentist simulation-based analysis showed any statistically significant preference over a simple power law. 

Nevertheless, a  number of mechanisms have been proposed \cite{Iqbal15} to deal with the possible suppression of the power spectrum on large scales and low multipoles.  
 In addition to being  an artifact of cosmic variance \cite{PlanckXX,Efstathiou03a}, large-scale power suppression could arise from changes in  the effective inflation-generating potential  \cite{Hazra14}, differing initial conditions at the beginning of inflation 
 \cite{Berera98,Contaldi03,Boyanovsky06,Powell07,Wang08,Broy15,Cicoli14,Das15,Mathews15}, the ISW effect \cite{Das14b}, effects of spatial curvature \cite{Efstathiou03b}, non-trivial topology \cite{Luminet03}, geometry \cite{Campanelli06,Campanelli07}, a violation of statistical anisotropies \cite{Hajian03}, effects of  a cosmological-constant type of dark energy during inflation \cite{Gordon04}, the bounce due to a  contracting phase to inflation \cite{Piao04,Liu13}, the production of primordial micro black-holes  \cite{Scardigli11}, hemispherical anisotropy and non-gaussianity \cite{McDonald14a,McDonald14b}, the scattering of the inflationary trajectory in multiple field inflation by a hidden feature in the isocurvature direction \cite{Wang15}, brane symmetry breaking in string theory \cite{Kitazawa14, Kitazawa15}, quantum entanglement in the M-theory landscape \cite{Holman08}, or loop quantum cosmology \cite{Barrau14}, etc.  
 

 In a previous work \cite{Mathews15},  we considered  another  
 possibility, i.e.~ that  the suppression of the power spectrum in the range $\ell = 10-30$ in particular could be due to  the resonant creation \cite{chung00,Mathews04} of Planck-scale fermions that
 couple to the inflaton field.  
 
   The present paper is an extension of that work.  Here, we propose  that both the suppression of the $\ell=2$ moment and the  suppression of the power spectrum in the range $\ell = 10-30$ could be explained from the resonant coupling to successive excitations  of a single  closed fermionic or bosonic superstring.  Indeed, both the apparent amplitude and the location of these features arise naturally in this picture.  There is also another possible string excitation for $\ell\approx 60$.   
   
   This result is significant in that accessing the mass scales of superstrings is impossible in the laboratory.  Indeed, the only place in Nature where such scales exist is during the first instants of cosmic expansion in the inflationary epoch.  Here we examine the possibility that, of the myriads of string excitations present in the birth of the universe out of the M-theory landscape, it may be that one string serendipitously made its presence known via a natural coupling to the inflaton field during the  $\sim 9$ $e$-folds visible on the sky.

 We emphasize, however, that the existence of such features in the CMB power spectrum from string theory is not unique. In\cite{Kitazawa14,Kitazawa15}  the suppression of the $\ell=2$ and the dip for $\ell =10 - 30$ were simultaneously  fit in a string-theory brane symmetry breaking mechanism.  In this case, however, the source of the features is due to the nature of the inflation-generating potential in string theory.  This mechanism splits boson and fermion excitations, leaving behind an exponential potential that is  too steep for the inflaton to emerge from the initial singularity while descending it. As a result, the scalar field generically "bounces against an exponential wall."  Just as in \cite{Hazra14}, this steepening  potential then introduces an infrared depression and  a pre-inflationary break in the power spectrum of scalar perturbations, reproducing the observed feature. 
 
In the present work, however, rather than to address the implications for the inflation-generating potential,  we consider the possibility of the resonant creation of  closed fermionic (or bosonic) superstrings with sequential excitations.  We also note that there may be a third marginally observable dip in the CMB power spectrum near $\ell \approx 60$.  Our goal is to demonstrate a proof or principle that it may be possible to identify string-like features in the CMB.  The goal here cannot be to provide the final formulation of a string theoretic explanation for deviations in the CMB power spectrum that is a fully realistic particle physics model.  This paper aims at cosmology not particle physics.  Hence we utilize a simple model that has some of the relevant coarse features of string theory.
\section{Resonant Particle Production during Inflation}
\label{sec: partprod}
The details of the  resonant particle creation paradigm during inflation have been explained in Refs.~\cite{chung00,Mathews04,Mathews15}.  Indeed, the idea was originally introduced \cite{Kofman94} as a means for reheating after inflation.
Since Ref.~\cite{chung00}, subsequent work \cite{Elgaroy03, Romano08, Barnaby09, Fedderke15}  has elaborated on the basic scheme into a model with coupling between two scalar fields.  
Here,  we summarize the essential features of a single fermion field coupled to the inflaton as a means to clarify the  physics of the possible dips in the CMB power spectrum.

In this minimal extension from the basic picture, the
inflaton $\phi$ is postulated to couple to particles
whose mass is of order the inflaton field value.  These particles
are then resonantly produced as the field obtains a critical value
during inflation.  If even a small fraction of the 
inflaton field is affected  in this way, it can produce an observable feature in
the primordial power spectrum. In particular, there can  be either an excess in the power spectrum as noted in \cite{chung00,Mathews04},
or a dip in the power spectrum as described in Ref.~\cite{Mathews15}. Such a dip offers important new clues to the trans-Planckian physics of the early universe.

In the simplest slow roll approximation \cite{starobinsky,Liddle,cmbinflate}, the generation of primordial density perturbations of  amplitude, $\delta_H(k)$ when crossing the Hubble radius is just,
\begin{eqnarray}
~~~~~~~~~~~~~~~~~~~~~~~~~~~~~~~~~~~~\delta_H(k) \approx {H^2 \over 5 \pi \dot \phi}~~,
\label{pert}
\end{eqnarray}
where $H$ is the expansion rate, and $\dot \phi$ is the rate of change of
the inflaton field when the comoving wave number $k$ crosses the
Hubble radius during inflation. The  resonant particle production
could, however,  affect the inflaton field such that  the conjugate momentum $\dot \phi$ is  altered.  
This could cause either an increase or a diminution in
$\delta_H(k)$ (the primordial power spectrum) for those wave numbers which
exit the horizon during the resonant particle production epoch.
In particular,  when $\dot{\phi}$ is accelerated  due to particle production,
it may deviate from the slow-roll condition.  In \cite{chung00}, however, this correction was analyzed
and found to be $<<20 \%$.  Hence, for our purposes we ignore this correction.

Here as in  \cite{chung00, Mathews04, Mathews15}, the effect of the resonant {\em fermionic} particle
production neglects  the non-adiabatic effects
on the modes outside of the horizon.  This leads to a dip-like
structure in the primordial power spectrum.  We caution, however, that in Ref.~in \cite{Elgaroy:2003hp} non-adiabatic effects  on the modes outside the horizon in the case of bosonic particle production 
were considered.  They deduced that the 
bosonic  primordial power spectrum is modified into a step-like structure
rather than a bump-like structure.  This would slightly modify the fit parameters.
For our purposes, however, we illustrate fermionic resonant particle production, but keep in mind that either a 
fermion or boson could produce the cosmological effects of interest here.

Hence, as in  \cite{Mathews15} we write the total Lagrangian density including the inflaton scalar field $\phi$, a  Dirac fermion field, and
the Yukawa interaction term as simply,
\begin{equation}
{\cal L}_{\rm tot} = \frac{1}{2}\partial_\mu \phi  \partial^\mu \phi - V(\phi)+ i \bar \psi\partial _{\mu}\gamma^\mu
\psi-m \bar \psi  \psi + N \lambda \phi \bar \psi  \psi ~~.
\end{equation}
For this Lagrangian, it is obvious that the effective fermion mass is:
\begin{equation} 
M(\phi) = m - N \lambda \phi~.
\end{equation}
  This vanishes
for a critical value of the inflaton field,
$\phi_* = m/N \lambda$.  Resonant fermion production
will then occur in a narrow range of the inflaton field amplitude
around $\phi = \phi_*$.


As in Refs.~\cite{chung00,Mathews04,Mathews15} we label the epoch at which particles are created
by an asterisk.  So, the cosmic scale factor is labeled $a_*$ at the
time $t_*$ at which resonant particle production occurs.  Considering
a small interval around this epoch, one can treat $H = H_*$ as
approximately constant (slow roll inflation).  The number density $n$
of particles can be taken \cite{chung00,Mathews04,Mathews15} as zero before $t_*$ and afterwards as $n =
n_*[a_*/a(t)]^{3}$.  The fermion vacuum expectation value can then be
written,
\begin{equation}
 \langle \bar \psi \psi \rangle = n_* \Theta (t-t_*) \exp{[-3 H_*(t-t_*)]} ~~.
 \end{equation}
where $\Theta$ is a step function.

Then following the derivation in \cite{chung00,Mathews04}, we can write the  modified equation of motion for the scalar field coupled to $\psi$:
\begin{equation}
\ddot \phi + 3 H \dot \phi = -V'(\phi) +  N \lambda \langle \bar \psi \psi \rangle ~~,
\end{equation}
where $V'(\phi) = dV/{d\phi}$.
The solution to this differential equation after particle creation $(t>t_*)$ is then similar to that derived in Refs.~\cite{chung00,Mathews04} but with a sign change for the coupling term, i.e.
\begin{eqnarray}
\dot \phi(t > t_*) &=& \dot \phi_* \exp{[-3H(t-t_*)]}\nonumber \\
& -& \frac{V'(\phi)_*}{3 H_*} \bigl[ 1 - \exp{[-3H(t-t_*)]}\bigr] \nonumber \\
&+&  N \lambda n_* (t-t_*) \exp{[-3 H_*(t-t_*)]} ~~.
\end{eqnarray}
The physical interpretation here is that the rate of change of the amplitude of the scalar field rapidly increases due to the coupling to particles created at the resonance 
$\phi = \phi_*$.  

Then, using Eq.~(\ref{pert})
for the fluctuation as it exits the horizon, and  constant $H \approx H_*$, one obtains  the perturbation in the primordial power spectrum as it exits the horizon:
\begin{equation}
\delta_H = \frac{[\delta_H(a)]_{N \lambda = 0}}{1 + \Theta (a - a_*)( N \lambda n_*/\vert \dot \phi_*\vert H_*) (a_*/a)^3 \ln{(a/a_*)}} ~~,
\label{deltahnew}
\end{equation}
where $\Theta(a-a_*)$ is the Heaviside step function.  It is clear in Eq.~(\ref{deltahnew}) that the power in the fluctuation of the inflaton field will abruptly diminish when the universe
grows to some critical scale factor $a_*$ at which time particles are resonantly created.

Using $k_*/k = a_*/a$, then
the perturbation spectrum Eq.~(\ref{deltahnew})
can be reduced \cite{Mathews04} to a simple 
two-parameter function.
\begin{equation}
\delta_H (k) = \frac{[\delta_H(a)]_{N \lambda = 0}}{1 + \Theta (k-k_*)A (k_*/k)^3 \ln{(k/k_*)}} ~~.
\label{perturb}
\end{equation}
where the amplitude  $A$ and characteristic wave number $k_*$  can be related to  the observed power spectrum from the approximate relation:
\begin{equation}
k_* \approx  \frac{ \ell_* }{ r_{lss}}~~, 
\end{equation}  
where $r_{lss} $ is the
comoving distance to the last scattering surface, taken here to be 13.8 Gpc \cite{PlanckXIII}.
For each resonance the values of $A$ and $k_*$ determined  from the CMB power spectrum
relate to the inflaton coupling $\lambda$ and fermion masses $m$ via Eqs.~(\ref{deltahnew}) and (\ref{perturb}). 
\begin{equation}
A  = |\dot{\phi}_*|^{-1} N \lambda
n_* H_*^{-1} ~~.
\end{equation}

 The connection between 
resonant particle creation and the CMB derives from the usual expansion  in spherical harmonics, $ \Delta T/T =\sum_l\sum_m
a_{lm}Y_{lm}(\theta,\phi)$ ($2 \le l<\infty$ and $-l \le m \le l$). 
The anisotropies are then described by the angular 
power spectrum, $C_l= \langle |a_{lm}|^2\rangle$, as 
a function of multipole number $l$.  One then merely
requires the conversion from perturbation spectrum $\delta_H (k)$
to angular power spectrum $C_l$.  This is easily 
accomplished using the {\it CAMB} code \cite{Camb}.
When converting to the angular power spectrum,
the amplitude of the narrow particle creation
feature  in $\delta_H(k)$ is spread over many values of $\ell$.
Hence, the particle creation features look like  broad dips in the power spectrum.




\section{Toroidal Compactification and the String Mass Spectrum}

\label{sec:toroidal}

 As a minimal step toward an analysis of  trans-plankian strings coupled to inflation  we consider the simplest compactified superstring.
The mass spectrum for  the simplest case of a closed bosonic string in 26 dimensions in which one of them is compactified into a circle \cite{Green,Polchinski} is:
\begin{equation}
M^2=\frac{n^2}{R^2}+\frac{w^2R^2}{{\alpha}'^2}+\frac{2}{{\alpha}'}(N+\tilde{N} - 2)~~.
\label{eq:1}
\end{equation}
Here, the integer $n$ labels the compact momentum eigenvalues. R is the radius of the compactified dimension, $w$ is the winding number describing the number of times the string wraps around the compactified dimension so that the second term gives the potential energy of the winding string. 
For the last term $N_{osc} \equiv (N+\tilde{N} - 2)$ counts the leftward moving and rightward oscillators along the  dimensions of  the string and the  zero point motion, where the oscillator number operators $N$ are 
\begin{equation}
N = \sum (\alpha_{-n}^{\mu}\alpha_{n\mu}+\alpha_{-n}\alpha_{n})~~.
\label{eq:2}
\end{equation}
\begin{equation}
\tilde{N} = \sum (-\tilde{\alpha}_{-n}^{\mu}\tilde{\alpha}_{n\mu}+\tilde{\alpha}_{-n}\tilde{\alpha}_{n})~~.
\label{eq:3}
\end{equation}
with,
\begin{equation}
N-\tilde{N}+nw = 0~~,
\label{eq:4}
\end{equation}
Note that for the $\alpha_n^\mu$ and $\tilde \alpha_n^\mu$, the index $\mu$ is over  the first 25-dimensions, while $\alpha_n$ and $\tilde \alpha_n$ refer to the compactified 25th dimension.

Eq.~(\ref{eq:1}) is a manifestation of the T-duality in string theory whereby for small compact dimensions string excitations are dominated by the momentum states of the compact dimension,  while for large dimensions  
the winding  states of the string become massive.   Moreover, the $R\rightarrow 0$ and $R \rightarrow \infty$ states are physically invariant in the mass spectrum, Eq.~(\ref{eq:1}).  That is, these states  are invariant under the coordinate transformation $R \rightarrow R' = \alpha'/R$ and $n \leftrightarrow w$.  Hence, in what follows states with different $n$ could either refer to momentum states  or different  winding numbers on the superstring. 

Although Eq.~(\ref{eq:1}) is for a bosonic string, we note that fermions are constructed from a combination of right going and left going modes on the string while imposing the appropriate (NS-R, R-NS) boundary conditions on a bosonic string.  Then, 
to obtain closed fermionic strings, the theory needs  to be realized in the SU(n) or SO(2n) group. We take $n=5$ M-theory. 
However, the same mass formula, Eq.~\ref({eq:1}) is  valid for an arbitrary compactification of fermionic strings as well as bosonic strings.  Although this is a very crude string theory, we identify two cases of cosmological interest.   

 In the limit of a fixed winding number and/or momentum state the string excitations can be identified with oscillations on the string.  Then
one can approximately write: 
 \begin{equation}
M^2\approx  \biggl(\frac{N_{osc} + \xi}{\alpha'}\biggr)  ~~,~~ {\rm Case~I}.
\label{eq:7}
\end{equation}
with 
\begin{equation}
\xi \equiv \alpha' \biggl(\frac{n}{ R}\biggr)^2 ~~.
\label{xi:eq}
\end{equation} 

The second case is that in which number of oscillations is fixed and $N - \tilde N = 0$.  Then the spectrum of momentum states on the string will be approximately 
\begin{equation}
M^2 \approx  \biggl(\frac{n^2 + \xi}{R^2}\biggr) ~~ ,~~{\rm Case~ II}.
\label{eq:6}
\end{equation}
with 
\begin{equation}
\xi = \frac{2R^2}{{\alpha}'}(N+\tilde{N}-2)
\label{case1}
\end{equation}
 For special circumstance of the ground state one has  $N=-\tilde{N} =1$.  
 
In principle, one could distinguish between these two cases if one could accurately determine the mass spectrum.  In the case of small $R$ and small $\xi$, the mass spectrum of momentum states should be regularly spaced, $M \propto n$.    On the other hand,  in the case of large R, the spacing of string mass states should be proportional to the square root of the number of oscillations $M \propto \sqrt{ N_{osc}}$.  Unfortunately, as noted below, the uncertainty in the mass spectrum is too large to distinguish which of these spectra best characterizes the deviations in the primordial power spectrum.

\subsection{String excitations and the CMB}
 In our previous paper \cite{Mathews15} we related the mass of the resonant particle to the scale $k^*$ and the number of $e$-folds ${\cal N_*}$ of inflation after the present associated scale left the horizon \cite{Liddle}.  This follows for any general monomial inflation 
 effective potential.
 That is, the resonance condition relates the mass $m$ to $\phi_*$ via,
 \begin{equation}
 m = N \lambda \phi_* ~~,
 \label{rescon}
 \end{equation}
 However, for a general monomial potential,
 \begin{equation}
V(\phi) = \Lambda_\phi m_{pl}^4 \biggl(\frac{\phi}{m_{pl}}\biggr)^\alpha~~,
\label{Vphi}
\end{equation}
there is an analytic solution for $\phi_*$ for a given scale in terms of the number of $e$-folds of inflation ${\cal N}_*$
\begin{equation}
\phi_* = \sqrt{2 \alpha {\cal N}_*} m_{pl}~~,
\end{equation}
where ${\cal N}_* $ is the number of $e$-folds of inflation corresponding to a given scale $k_*$, 
\begin{equation}
{\cal N}_* = \frac{1}{m_{\rm pl}^2} \int_{\phi_{end}}^{\phi_*} \frac{V(\phi)}{V'(\phi)} d\phi~= {\cal N} - \ln{(k_*/k_H)} ~~,
\end{equation}
where  $\phi_{end}$ is the value of the scalar field at the end of inflation, ${\cal N}$ is the total number of $e$-folds of inflation and the Hubble scale is $k_H = h/2997.3  = 0.000227$ Mpc$^{-1}$ (for $h = 0.68$) \cite{PlanckXIII}.
 
 So, for the compactified superstrings we can write 
 \begin{equation}
 M =  N \lambda \phi_*  = N \lambda \sqrt{2 \alpha} \sqrt{{\cal N} - \ln{(k_*/k_H)} }~m_{pl}~,
\label{Meq}
 \end{equation}
and  we can write the mass corresponding to a given multipole on the sky
 \begin{equation} 
 M(\ell_*)^2 \propto ({\cal N} - \ln{(k_*/k_H)})~~.
 \end{equation}
 
 Next, we make  the simplifying assumption that the resonant states in the spectrum  differ only in the number of excitations on the string.
Then the coupling to the inflaton field $\lambda$ is the same, along with the number of degenerate fermion states $N$ at a given mass.  We also keep the same normalization of the mass scale 
$\alpha'$.

Then if we take ${\cal N} = 50$, we can write for the ratio of the quadrupole ($\ell^* = 2$) suppression resonance to the $\ell^* \approx 20$ resonance:
\begin{equation}
\frac{M^2(\ell^*=2)}{M^2({\ell^*=20})}  \equiv {\cal R}_{+1} \approx \frac{{\cal N} - \ln{(k_*(n+1)/k_H)}}{{\cal N} - \ln{(k_*(n)/k_H)}} ~~.
\label{ratio2-20}
\end{equation}
Similarly for the higher multipoles we can define:
\begin{equation}
\frac{M^2(\ell^*=20)}{M^2({\ell^*=60})}  \equiv {\cal R}_{-1} \approx \frac{{\cal N} - \ln{(k_*(n)/k_H)}}{{\cal N} - \ln{(k_*(n-1)/k_H)}} ~~.
\label{ratio20-60}
\end{equation}
Hence, from fits to the CMB, one can deduce the ratio of excited states on the superstring in this simple model.

\section{ $\chi^2$ Fit to the CMB}
 We have made a straightforward $\chi^2$ minimization to fit the 
TT CMB {\it Planck}  power spectrum  \cite{PlanckXIII}  for the $\ell^* =  2$ and $\ell^* \approx 20$ resonances.  We also searched for a possible third dip in the spectrum.   For simplicity and speed  we fixed all cosmological parameters at the values deduced by {\it Planck} \cite{PlanckXIII} and only searched over a single  amplitude. We note, however, that this straightforward fit does not take into account the off-diagonal $\ell-\ell$ terms. This approximation is reasonable in the TT case where these terms can be negligible (however, they are not exactly zero because of the the presence of a Galactic mask). On the other hand, in the polarization case (EE power spectrum) those terms are expected to be much more important. This is addressed in the following section where we make a separate Markov Chain Monte-Carlo fit to the combined TT, TE, and EE power spectrum in which the full correlation matrix is incorporated.

From this simple $\chi^2$ fit we  deduce the following resonance parameters:

$$\ell \approx 2, ~A = 1.7\pm 1.5, ~k_* (n+1)= 0.0004 \pm 0.0003 ~h~{\rm Mpc}^{-1}$$
$$\ell \approx 20,~A = 1.7\pm 1.5 ,~k_*(n) = 0.0015 \pm 0.0005  ~h~{\rm Mpc}^{-1}$$
$$\ell \approx  60,~A = 1.7 \pm 1.5,~k_*(n-1) = 0.005 \pm 0.004  ~h~{\rm Mpc}^{-1}$$

\begin{figure}
\includegraphics[width=0.49\textwidth]{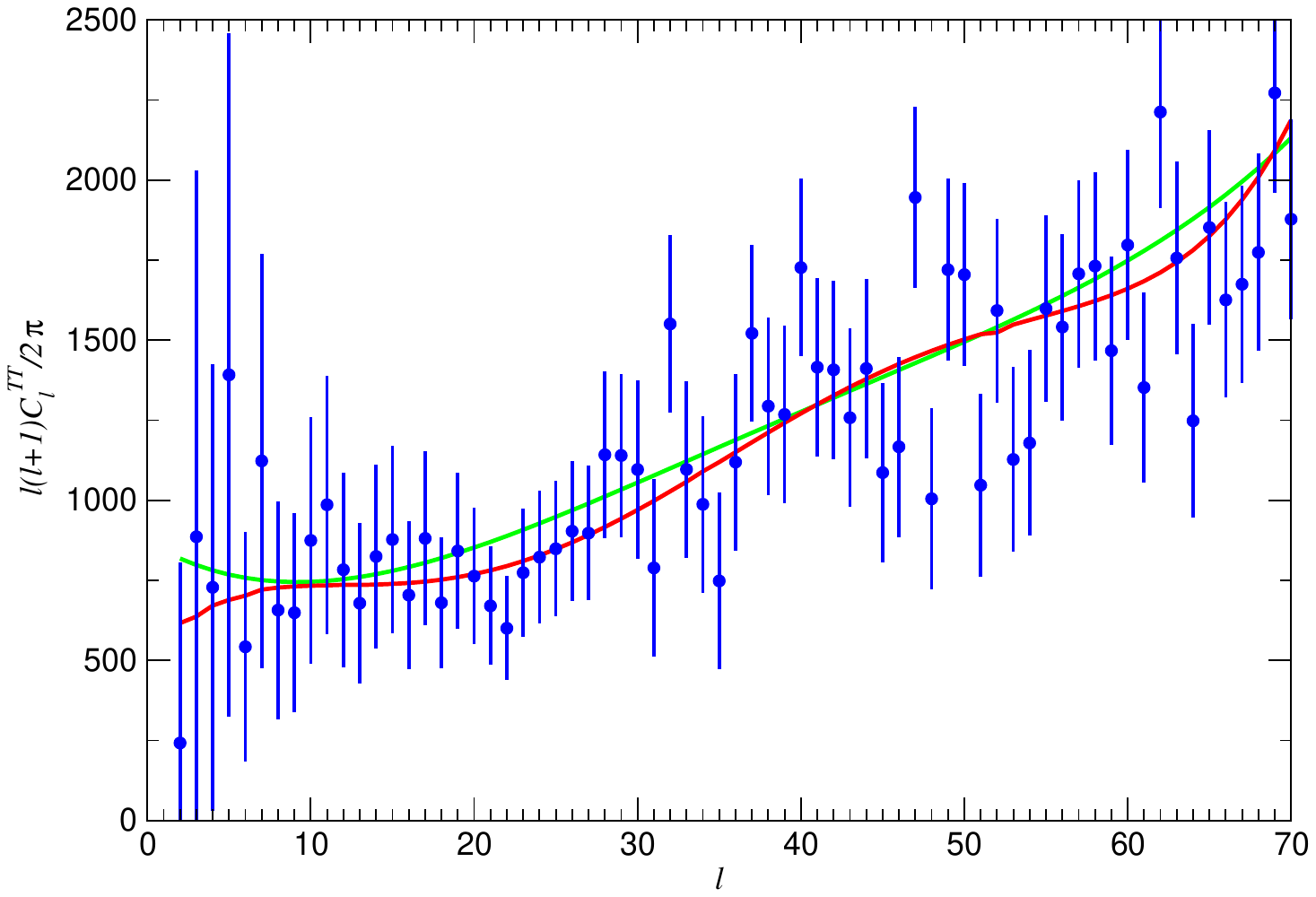} 
\caption{(Color online) The fit (red line) to $\ell \approx 2$, $\ell \approx 20$ and  $\ell \approx 60$ suppression of the {\it TT}  CMB power spectrum as described in the text. Points with error bars are from the  {\it Planck} Data Release \cite{PlanckXIII}.  The green line shows the best standard $\Lambda$CDM power-law fit to the {\it Planck} CMB power spectrum}
\label{fig:1}
\end{figure}

Figure \ref{fig:1} illustrates  the best  fit to the {\it TT} CMB power spectrum that includes both the ${\ell } \approx  2$,  ${\ell } \approx  20$ and ${\ell } \approx  60$ suppression of the CMB.

It is obvious from Figure \ref{fig:1} that that the evidence for this fit is statistically weak due to the large errors in the data.  Indeed, the  total reduction in $\chi^2$  is $\Delta \chi^2 = -9$  for a fit with an addition of 3 degrees of freedom, i.e. the amplitude $A$  and two independent values for $k_*$.  

\begin{figure}
\includegraphics[width=0.49\textwidth]{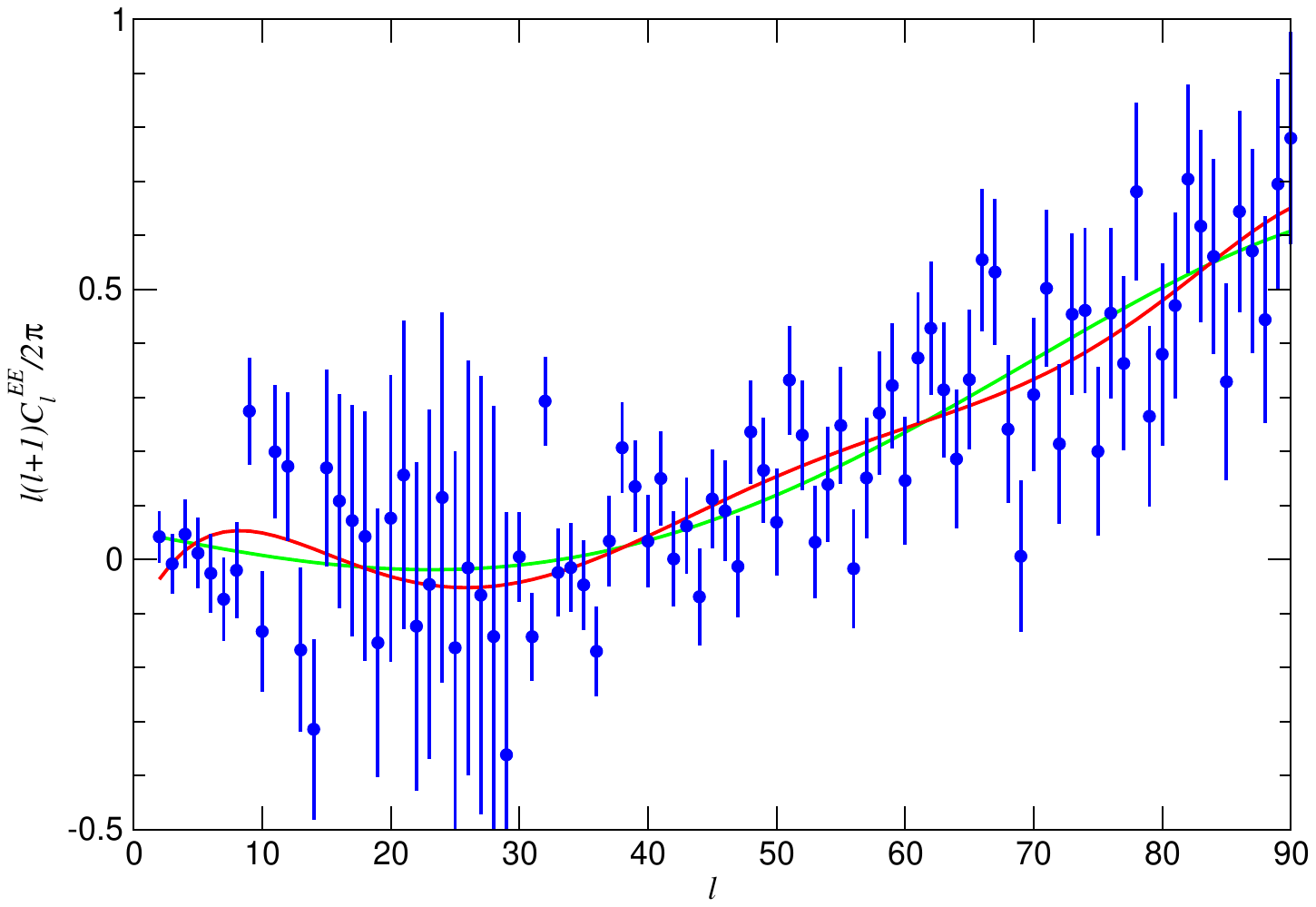} 
\caption{Same as Figure \ref{fig:1} but in this case the lines are the derived  {\it EE} CMB power spectrum based upon the fits to the {\it TT} power spectrum shown in Figure \ref{fig:1}.}
\label{fig:2}
\end{figure}

Figure \ref{fig:2} similarly illustrates  the derived  {\it EE} CMB power spectrum based upon the fits to the {\it TT} power spectrum shown in Figure \ref{fig:1}.  Although this fit is not optimized, and the uncertainty in the data is large,  there is a reduction in total $\chi^2$ by $\Delta \chi^2 = -5$ for the line with resonant superstring creation.  Hence, the {\it EE} spectrum is at least consistent with this paradigm and in fact slightly favors it.

Under the assumption that the model errors are independent and obey a normal distribution, 
 then the Bayesian information criterion (BIC)  can be written \cite{Mathews15} in terms of $\Delta \chi^2$ as 
$\Delta$BIC$ \approx \Delta \chi^2  + (p \cdot \ln{n})$, where $p$ is the number of parameters in the test and $n$ is the number of points in the observed data. When selecting the best  model,  the lowest BIC is preferred since the BIC is an increasing function of both the error variance and  the number of new degrees of freedom $p$. In other
words, the unexplained variation in the dependent variable and the number of explanatory variables increase the value of BIC. Hence, a negative $\Delta$BIC implies either fewer explanatory variables, a better fit, or both. For the $\approx 140$ data points in the range of the fits of Figure \ref{fig:1} plus \ref{fig:2},  the inferred total improvement is $\Delta \chi^2 = -14$ with the  introduction of 3 new parameters.   This corresponds to a $\Delta$BIC$ = +0.8$. Generally, $\Delta$BIC$>2$ is required to be considered evidence against  a particular model.  Hence, one must conclude that  although the fit including the superstring resonances produces an improvement in $\chi^2$, it is  statistically equivalent to the simple power-law fit.
 Nevertheless, it is worthwhile to examine the possible physical meaning of the deduced parameters.
 Based upon our fit to the three possible resonances in the CMB, we deduce from Eqs.~(\ref{ratio2-20}) and (\ref{ratio20-60}) the following ratio of excited states:
$ \frac{M^2(\ell^*=2)}{M^2({\ell=20})}  \equiv {\cal R}_{+1}  = 1.024 \pm 0.050$.  Surprisingly, we also obtain
$\frac{M^2(\ell^*=20)}{M^2({\ell=60})}  \equiv {\cal R}_{-1}  = 1.024 \pm 0.030$.  Hence we deduce that there is a regular spacing in the mass spectrum of these three states.\\*

As an illustration of how the results of the $\chi^2$ fit might relate to string parameters let us consider the simplest possible example.
For Case I simple oscillations on a string  in the limit of large $R$ then one simply has
 \begin{equation}
{\cal R}_{+1} = \frac{(N_{osc}+1)}{N_{osc}} ~~.
\label{R+eq}
\end{equation}
From which one could deduce
\begin{equation}
N_{osc} = \frac{1}{{\cal R}_{+1}-1}~~.
\end{equation}
For ${\cal R}_{+1}  = 1.024 \pm 0.050$ one could then  deduce $N_{osc} = 42^{+\infty}_{-28}$ for the number of oscillations on 
the compactified fermionic string.  Obviously, the uncertainty is quite large.  Nevertheless, this illustrates the possibility to identify the string excitation.

One can also place some constraint on the mass and coupling constant.
The amplitude  $A$
can be related directly to the coupling constant $\lambda$ 
using the 	following approximation
 for the
particle production Bogoliubov coefficient \cite{chung00,birrellanddavies,Kofman:1997yn,Chung:1998bt}
\begin{equation}
|\beta_k|^2 \approx \exp\left( \frac{-\pi k^2}{a_*^2  \lambda |\dot \phi_*| }\right).
\end{equation}

Then,
\begin{equation}
\label{eq:nstar}
n_* = \frac{2}{\pi^2}\int_0^\infty dk_p \, k_p^2 \, 
|\beta_k|^2 =
\frac{N \lambda^{3/2}}{2\pi^3}
|\dot{\phi}_*|^{3/2}~~ .
\end{equation}
This gives, 
\begin{eqnarray}
A & = & \frac{ N \lambda^{5/2}}{2 \pi^3}
\frac{\sqrt{|\dot{\phi}_*|}}{H_*}\\
& \approx & \frac{N \lambda^{5/2}}{2 \sqrt{5} \pi^{7/2}}
\frac{1}{\sqrt{\delta_H(k_*)|_{\lambda=0}}}~~.
\label{eq:alamrelation}
\end{eqnarray}
 where we have used the usual approximation for the primordial slow
roll inflationary spectrum \cite{Liddle,cmbinflate}.  


Now, given that
the CMB normalization requires that\
$\delta_H(k)|_{\lambda=0}\sim
10^{-5}$, we  have 
\begin{equation}
A \sim 1.3 N  \lambda^{5/2}.
\label{eq:avnl}
\end{equation}
Hence, for the maximum likelihood value of $A \sim 1.7 \pm 1.5$, we have 
\begin{equation}
\lambda  \approx \frac{(1.1 \pm 1.0)}{N^{2/5}} ~~.  
\label{Nconst}
\end{equation}
 The fermion particle mass $m$ can then be deduced from the resonance condition, $m = N \lambda \phi_*$.  

From Eq.~(\ref{Nconst}) then we have
$m \approx   \phi_*/\lambda^{3/2}$.
For the $\ell \approx 20$ ($k_* = 0.0015 \pm 0.0005 ~h~{\rm Mpc}^{-1}$) resonance, and $k_H = a_0 H_0 = (h/2997.9)$ Mpc$^{-1} \sim 0.0002,$ we have ${\cal N} - {\cal N}_*  = ln{(k_H/k_* )} <1$.  Typically one expects  ${\cal N}(k_*) \sim  {\cal N} \sim 50 - 60$.   

We can then apply the resonance condition  [Eq.~(\ref{rescon})] to deduce the approximate range  of masses for the string excitations.
Monomial potentials [Eq.~(\ref{Vphi})] with $\alpha = 2/3$  or $\alpha =1$ correspond to the lowest order approximation to the string theory axion monodromy inflation potential \cite{Silverstein08, McAllister10}.  Moreover, the limits on the tensor to scalar ratio from the {\it Planck} analysis \cite{PlanckXX}  are more consistent with  $\alpha = 2/3$ or $1$.   If we fix the value of $A = 1.7$, then from the range of 50-60 $e$-folds we would have $\phi_* = (8-9) ~m_{pl}$ for $\alpha = 2/3$ or $\phi_* = (10-11) ~m_{pl}$ for $\alpha = 1$.  
Hence,  we have roughly the constraint,  
\begin{equation}
m \sim  (8-11)  ~\frac{m_{\rm pl}}{\lambda^{3/2}}~~.
\label{masseq}
\end{equation}
We note, however, that if the uncertainty in the normalization parameter $A$ is taken into account, this range increases.  This illustration is simply meant to demonstrate that the mass of the string excitation can  be determined once the coupling constant is known.  

To find the coupling constant via Eq.~(\ref{Nconst}), one must know the degeneracy of the string states.  
However, the degeneracy of string states can be enormous, and is dependent upon a detailed model which is beyond the scope of this paper. 
For our purpose it is sufficient that the degeneracy is large, implying a small coupling consistent with our application of this simple resonant coupling model.  

\section{MCMC fit to the CMB}

The statistical significance of the $\chi^2$ fit is marginal.  However, as demonstrated above it could indicate some physical insight into the nature of the stringy landscape out of which the universe inflated.  As a next step in the analysis we also performed an independent multi-dimensional Markov Chain Monte Carlo (MCMC) fit to the  Planck 2015 \cite{PlanckXX} TT, TE,  EE power spectra \cite{PlanckXX}.  These fits are based upon the the publicly available CosmoMC \cite{cosmomc}.   This analysis complements the straight forward analysis carried out above and leads to  a somewhat different possible physical implication. 

We utilized the Planck results \cite{PlanckXX} along with the results of the $\chi^2$ analysis as priors.  We then sample over the standard cosmological parameters $(\Omega_b h^2,\Omega_c h^2,$ $\tau, n_s, \theta, log(10^{10}A_s))$ where, $\Omega_{b}h^2$ and $\Omega_{b}h^2$ (with $h$ related to the present Hubble parameter) represent baryon and dark matter densities respectively, while $n_s$ and $\tau$ are the scalar spectral index and the reionization optical depth respectively. For the other two parameters, $\theta$ is the angle subtended by the sound horizon at recombination and $log(10^{10}A_s)$ is the logarithmic amplitude of the primordial perturbations.   We also sample over  the  resonance parameters, $(p_1, p_2, q_1, q_2, r_1, r_2)$ where  $p_1, q_1, r_1$ are the amplitudes of the resonances and $p_2, q_2,r_2$ are the respective resonance locations $l_*$.
We considered cases both with the amplitude of the resonance dips $A_i =p_1,q_1, r_1$  fixed at a common value and with the amplitudes allowed to vary from one resonance to the next.  

\begin{figure}
\includegraphics[width=0.5 \textwidth,height=17cm]{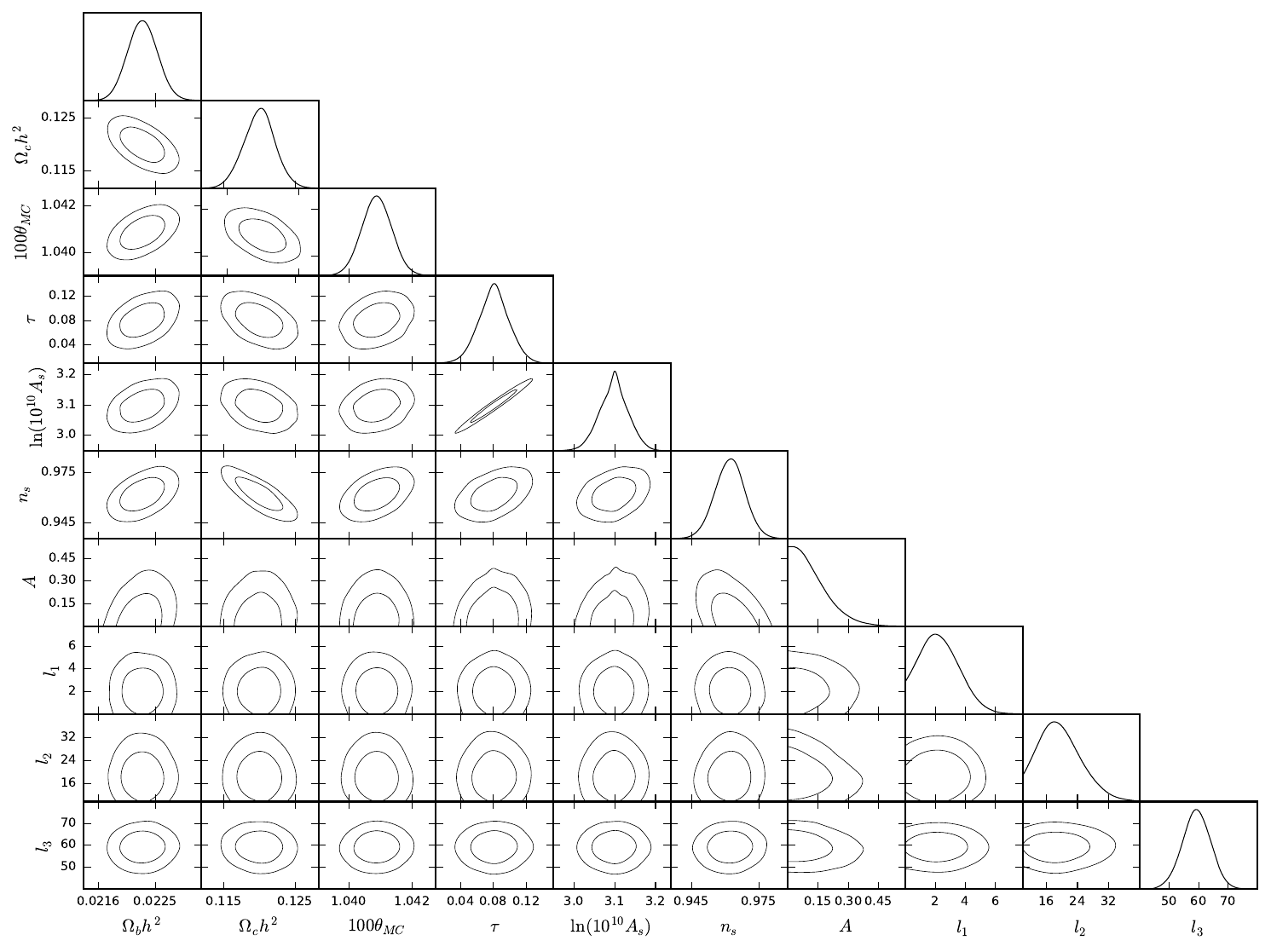} 
\caption{Contours for the  68$\%$ and 95$\%$ confidence limits for the resonance parameters  $A_*  = p_1= q_1 = r_1$ and multipoles $\ell_* = p_2, q_2, r_2$ for the three dips in the CMB power spectrum. }
\label{fig:3}
\end{figure}

Figure \ref{fig:3} illustrates  contours of marginalized probability densities for the cosmological and resonance parameters for the case in which the 3 amplitudes are at a fixed single value $A = p_1 = q_1 =r_1$ for the three resonances at $l_* = p_2, q_2$, and $r_2$ respectively.   This plot confirms that there are no significant correlations among parameters except for the familiar one between $N_s$ and $\Omega_c h^2$, and a very slight correlation between $A$ and $n_s$. There is also a striking result compared to the $\chi^2$ analysis that the amplitude $A$ for the resultant fit is diminished by an order of magnitude to  $A \le 0.16$.  The reason for this can be traced to to the fact that statistical power around $\ell=60$ is much larger than at $\ell=20$ or $\ell=2$. Hence, if we impose  a common amplitude $A$, then data around $\ell=60$ do not allow for a large amplitude.  

\begin{figure}
\includegraphics[width=0.5\textwidth, height=15cm]{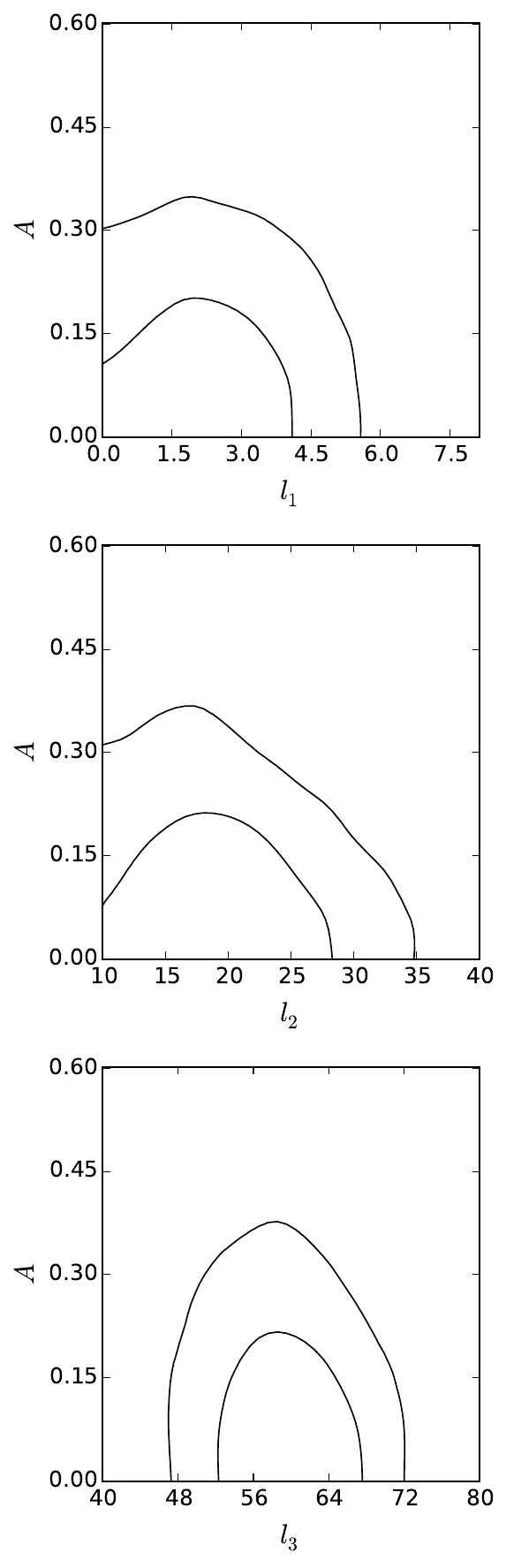} 
\caption{Contours for the  68$\%$ and 95$\%$ confidence limits for the resonance parameters  $A  = p_1, q_1, r_1$ and multipole $\ell_{1,2,3} \equiv p_2, q_2, r_2$ for the three dips in the CMB power spectrum. }
\label{fig:4}
\end{figure}

\begin{figure}
\includegraphics[width=0.5\textwidth, height=17cm]{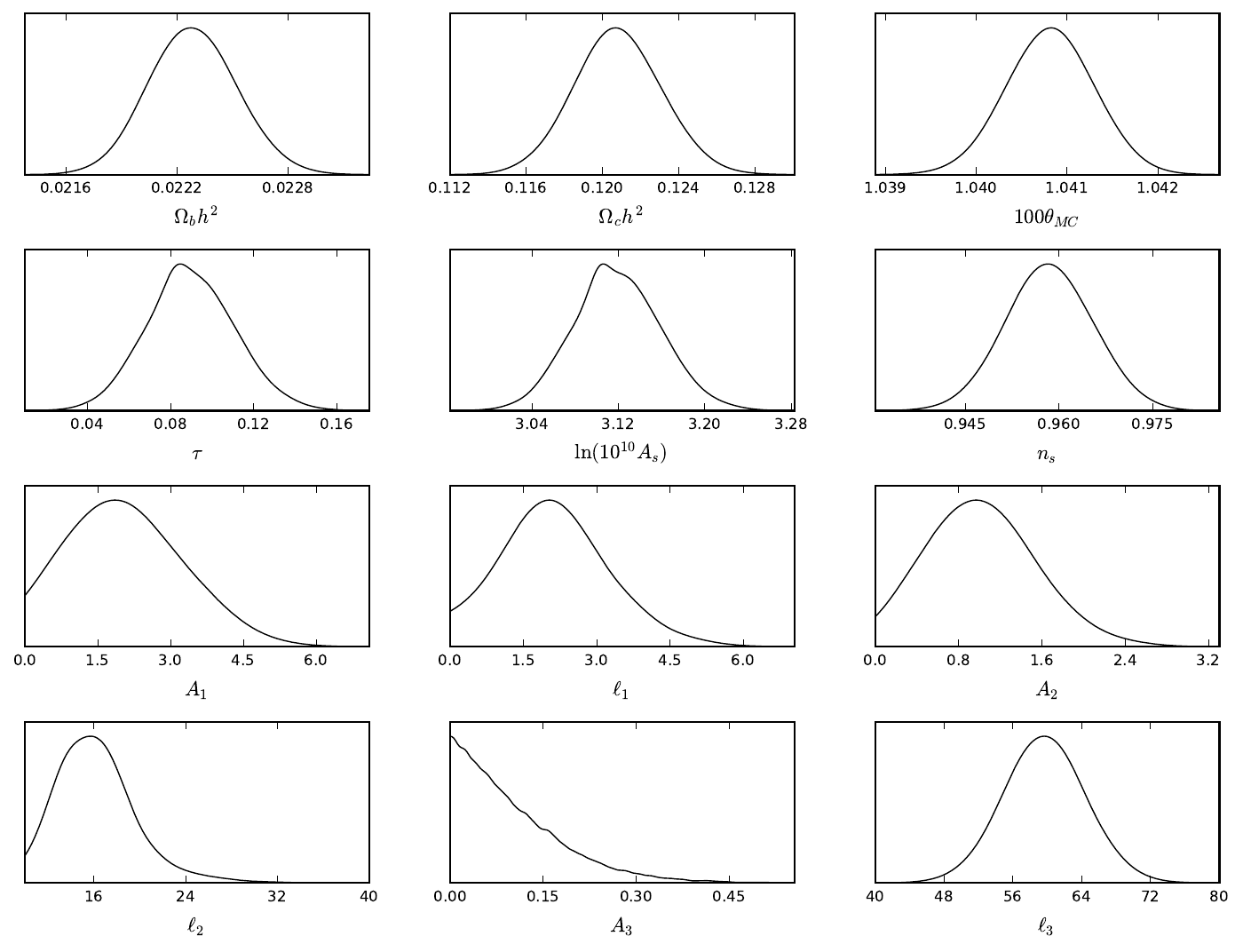} 
\caption{Marginalized likelihood functions of the resonance parameters and some cosmological observables as labeled. The resonance parameters are the same as in Figure \ref{fig:3}. }
\label{fig:5}
\end{figure}

Hence, it is illustrative to consider the case in which all three amplitudes are allowed to vary.  Figure \ref{fig:4} shows contours of marginalized probability densities for fits to the TT power spectrum for the case in which the 3 amplitudes, $A_i = p_1$, $q_1$, and $r_1$ are allowed to vary independently for the three resonances at $l_* = p_2, q_2$, and $r_2$, respectively. Similarly, Figure \ref{fig:5} shows the likelihood functions both for the resonances and the cosmological parameters for this case.
 
 Figures \ref{fig:4} and \ref{fig:5} show that unlike straightforward $\chi^2$ analysis, at best only an upper limit to  the amplitude  for the $\ell = 60$ resonance can be estimated. The best-fit values of the multipoles ($\ell$) representing the three dips are respectively $\ell = 1.7 \pm 4, 16 \pm 3$ and $60 \pm 4$.  However, the amplitude for the  3 resonances differ significantly, i.e. $p_1 = 1.7 \pm 4$, $q_1 = 1.0 \pm 6,$ and $r_1 \le 0.1$. 

If we take these amplitudes seriously, then there could be a physical interpretation.  Since $A \propto  N \lambda^{2/3}$,  where $N$ is the degeneracy, then we might be seeing a progression of the degeneracy of the string (for $\lambda$ fixed).  For example, If $N \propto N_{\rm osc}$ from Eq.~\ref{eq:1} where $N_{\rm osc}$ is the number of oscillations and the $\ell = 60$ resonance represents a zero point with $N = 0$ then the amplitude for the $\ell = 60$ resonance would be small while  the  resonance with $\ell \approx 20$ would have, $N_{\rm osc} = 1$, and the  resonance with $\ell \approx 2$. would have, $N_{\rm osc} = 2$.   This implies that the   ratio of the amplitude for the $\ell \approx 2$ resonance to the $\ell \approx 20$ should be about $A(\ell=2)/A(\ell = 20) = 2$.  This is the progression we see in the MCMC analysis.
Obviously, there is much uncertainty remaining in the analysis and the interpretation.  Our goal here is only to illustrate the  possibility to uncover the physical properties of a superstring resonantly coupled with the inflaton during inflation.

\section{Conclusion}

We have analyzed  dips in the {\it Planck} \cite{PlanckXX} CMB power spectrum at  $\ell \approx  2, 20$ and $\ell \approx 60$ as possible evidence for successive excitations of a superstring resonantly coupled with the inflaton during inflation. In a simple $\chi^2$ analysis the  best fit to these features implies dips in the primordial power spectrum with an amplitude of $A \approx  1.7 \pm 1.5$ corresponding to $\sim 40$ oscillations on the string.  An MCMC analysis, however, prefers a fit with  significant changes in the amplitude from one resonance to the next.  In a simplified string model this is suggestive of what could be expected for the first few oscillation states on a superstring. Although of marginal statistical significance, we suggested that these results are consistent with a simplified model for the resonant creation of successive excitations on a toroidal compactified  superstring during inflation. 
For  string-theory motivated axion monodromy  inflation potentials consistent with the {\it Planck} tensor-to-scalar ratio,  these features  would correspond to the resonant creation
of successive superstring momentum (or winding) states or oscillations with a large trans-Plankian mass. 
 
Obviously this simple phenomenological analysis should be done in the context of a more realistic string theory.  Also, there is a need for more precise determinations of deviations of the
CMB  power spectrum particularly  in the range of $\ell = 2-100$, although this may ultimately be limited by the cosmic variance.
Nevertheless, in spite of these caveats, we conclude that if the present
analysis is correct, this may be  the first  hints at observational
evidence of successive excitations of a superstring present at the Planck scale.  

Indeed, one
expects a plethora of superstring excitations to be  present when the universe exited from the M-theory landscape. Perhaps, the presently observed CMB power
spectrum contains the first suggestion that one of those many ambient  superstrings may have 
 coupled to the inflaton field during the $\sim$9 $e$-folds of inflation visible on the horizon, thereby leaving  behind a relic signature of its existence.

\begin{acknowledgements}
Work at the University of Notre Dame is supported
by the U.S. Department of Energy under 
Nuclear Theory Grant DE-FG02-95-ER40934.
Work at NAOJ was supported in part by Grants-in-Aid for Scientific Research of JSPS (15H03665, 17K05459). Work at Nagoya University supported by JSPS research grant number 24340048. MRG wants to thank N.~Kumar for valuable suggestions. KI wants to acknowledge JSPS research grant numbers 18K03616 and 15H05890.
\end{acknowledgements}


\begin{thebibliography}{9}

%
\bibitem{Martin01} J. Martin and R. H. Brandenberger, Phys. Rev. {\bf D63}, 123501 (2001)
%
\bibitem{vafaei} 
Vafaei Sadr, A.; Movahed, S. M. S.; Farhang, M.; Ringeval, C.; Bouchet, F. R., Mon.Not.Roy.Astron.Soc. 475 (2018) no.1, 1010-1022 .
%
\bibitem{fursaev} D.V. Fursaev, Phys. Rev. D 96, 104005 (2017).
%
\bibitem{eiguren} Asier Lopez-Eiguren, Joanes Lizarraga, Mark Hindmarsh, Jon Urrestilla, JCAP 1707 (2017) no.07, 026.
%
\bibitem{jellis} Ellis, John, Int. J. Mod. Phys. D, Volume 25, Issue 14, id. 1630027.
%
\bibitem{sinha} Gordon Kane, Kuver Sinha, Scott Watson, Int.J.Mod.Phys. D24 (2015) no.08, 1530022.
%
\bibitem{westphal} Alexander Westphal, Int.J.Mod.Phys. A30 (2015) no.09, 1530024.
%
\bibitem{mrgkahler}S.~Bhattacharya, K.~Dutta, M.~R.~Gangopadhyay, A.~Maharana, \ Phys.\ Rev.\ D97 (2018) 123533.
%
\bibitem{chernoff} David F. Chernoff, S.-H. Henry Tye, Int.J.Mod.Phys. D24 (2015) no.03, 1530010 
%
\bibitem{Mathews15} G.~J.~Mathews, M. R. Gangopadhyay, K. Ichiki, and T. Kajino, Phys. Rev. D92, 123519 (2015).
%
\bibitem{starobinsky} A.~A.~Starobinsky, Sov.~Astron.~Lett.~4, 83 (1978).
%
\bibitem{Liddle}
A. R. Liddle and D. H. Lyth, {\it Cosmological Inflation and Large Scale Structure}, (Cambridge University Press: Cambridge, UK), (1998).
%
\bibitem{cmbinflate}
E. W. Kolb and M. S. Turner, 
{\it The Early Universe}, (Addison-Wesley, Menlo Park, Ca., 1990).
%
\bibitem{PlanckXIII} {\it Planck} Collaboration, Astron. \& Astrophys. {\bf 594} A13 (2016).
%
\bibitem{PlanckXX} {\it Planck XX} Collaboration, Astron. \& Astrophys. {\bf 594} (2015) A20 (2016).
%
\bibitem{WMAP9} G. Hinshaw,  et al. ({\it WMAP Collaboration})    Astrophys. J. Suppl. Ser., {\bf 208}, 19 (2013).
%
\bibitem{Iqbal15} A. Iqbal, J. Prasad, T. Souradeep, and M. A. Malik,  JCAP, {\bf 06}, 014 (2015).
%
\bibitem{Efstathiou03a} G. Efstathiou,  Mon. Not. R. Astron. Soc. {\bf 346} L26, (2003).
%
\bibitem{Hazra14} D. K. Hazra, A. Shafieloo, G. F. Smoot, and A. A. Starobinsky, JCAP, {\bf 08}, 048 (2014).
%
%
\bibitem{Berera98} A. Berera, L.-Z. Fang, and G. Hinshaw, Phys. Rev. D {\bf 57}  2207, (1998).
%
\bibitem{Contaldi03}C. R. Contaldi, M. Peloso, L. Kofman, and A. Linde, JCAP {\bf 7}   2, (2003).
%
\bibitem{Green} Green, Scwartz, Witten {\it Superstring Theory }(Vol. 1) , (Cambridge University Press, Cambridge 1987).

\bibitem{Polchinski} J. Polchinski,  {\it String Theory} (Vol. 1), (Cambridge University Press, Cambridge 2005).

\bibitem{Boyanovsky06}D. Boyanovsky, H. J. de Vega, and N. G. Sanchez, Phys. Rev. D {\bf 74} 123006, ( 2006).
%
\bibitem{Powell07} B. A. Powell and W. H. Kinney, Phys. Rev. D {\bf 76}  063512, (2007).
%
\bibitem{Wang08} I.-Chin Wang and K.-W. Ng, Phys. Rev. D {\bf 77}  083501, (2008).
%
\bibitem{Broy15} B. J. Broy, D. Roest, and A. Westphal,  Phys. Rev.  D {\bf 91}, 023514, (2015) .
%
\bibitem{Cicoli14} M. Cicoli, S. Downes, B. Dutta, F. G. Pedro, and A. Westphal,  JCAP {\bf 12}  30, (2014).
%
\bibitem{Das15} S. Das, G. Goswami, J. Prasad, and R. Rangarajan, JCAP, {\bf 06}, 01 (2015).
%
\bibitem{Das14b} S. Das and T. Souradeep,  JCAP {\bf 2}, 2,  (2014).
%
\bibitem{Efstathiou03b} G. Efstathiou, Mon. Not. R. Astron. Soc. {\bf 343} L95 (2003).
%
\bibitem{Luminet03}J.-P. Luminet, J. R. Weeks, A. Riazuelo, R. Lehoucq, and J.-P. Uzan, Nature, {\bf 425} 593 (2003).
%
\bibitem{Campanelli06} L. Campanelli, P. Cea, and L. Tedesco, Phys. Rev. Lett. {\bf  97}  131302, (2006).
%
\bibitem{Campanelli07} L. Campanelli, P. Cea, and L. Tedesco, Phys. Rev. D {\bf 76} 063007, (2007).
%
\bibitem{Hajian03} A. Hajian and T. Souradeep,  Astrophys. J. Lett. {\bf 597} L5, (2003).
%
\bibitem{Gordon04} C. Gordon and W. Hu, Phys. Rev. D {\bf 70}  083003, (2004).
%
\bibitem{Piao04} Y.-S. Piao, B. Feng, and X. Zhang, Phys. Rev. D {\bf 69}  103520, (2004).
%
\bibitem{Liu13} Z.-G. Liu, Z.-K. Guo, and Y.-S. Piao,  Phys. Rev. D {\bf 88}  063539, (2013).
%
\bibitem{Scardigli11} F. Scardigli, C. Gruber, and P. Chen, Phys. Rev. D {\bf 83}  063507 (2011).
%
\bibitem{McDonald14a} J. McDonald, Phys. Rev. D {\bf 89}  127303, (2014).
%
\bibitem{McDonald14b} J. McDonald,  JCAP {\bf 11} 012, (2014).
%
\bibitem{Wang15} Y. Wang  and Y.-Z. Ma,  eprint arXiv:1501.00282v1 (2015).
%
\bibitem{Kitazawa14} N. Kitazawa and A. Sagnotti, EPJ Web of Conferences {\bf 95}, 03031 (2015).
%
\bibitem{Kitazawa15} N. Kitazawa and A. Sagnotti, Mod. Phys. Lett. A {\bf  30}, 1550137 (2015). 
%
\bibitem{Holman08} R. Holman, L. Mersini-Houghton, and T. Takahashi, Phys. Rev. D {\bf 77}, 063511 (2008).
%
\bibitem{Barrau14} A. Barrau, T. Cailleteau, J. Grain, and J. Mielczarek,  Classical and Quantum Gravity {\bf 31}  053001 (2014).
%
\bibitem{cosmomc} A.~Lewis, S.~Bridle,\ Phys.\ Rev. \ D66 (2002).
%
%
\bibitem{chung00} D. J. H. Chung, E. W. Kolb, A. Riotto, and I. I. Tkachev, Phys. Rev. D {\bf 62}, 043508 (2000).
%
\bibitem{Mathews04}G. J. Mathews, D.  Chung, K. Ichiki, T. Kajino, and M. Orito,  Phys. Rev. {\bf D70},  083505 (2004). 
%
\bibitem{Elgaroy:2003hp} O.~Elgaroy, S.~Hannestad and T.~Haugboelle,
JCAP {\bf 0309}, 008 (2003).
%
\bibitem{cbi1}
B. S. Mason,   et al. ({\it CBI} Collaboration), Astrophys. J.,  591, 540  (2003).
%
\bibitem{cbi2}
T. J. Pearson, et al. ({\it CBI} Collaboration), Astrophys. J.,  591, 556  (2003).
%
\bibitem{cbi3} A. C. S. Readhead, et al., Astrophys. J. in press, (2004).
%
\bibitem{acbar}
C. L. Kuo et al., ({\it ACBAR} Collaboration),  Astrophys. J., {\bf 600}, 32 (2004).
%
\bibitem{BIMA}
K. S. Dawson, et al., {\it BIMA} Collaboration),  Astrophys. J., {\bf 581}, 86 (2002).
%
\bibitem{VSA}
K. Grainge, et al., {\it VSA} Collaboration), Mon. Not. R. Astron. Soc., {\bf 341}, L23,   (2003).
%
\bibitem{WMAP1}
C. L. Bennett, et al. ({\it WMAP} Collaboration), Astrophys. J., Suppl., {\bf 148}, 99 (2003);
 D. L. Spergel, et al., Astrophys. J. Suppl., {\bf 148}, 175 (2003).
%
%
%
\bibitem{Kofman94} L. Kofman, A. D. Linde, and A. A. Starobinsky, Phys. Rev. Lett. {\bf 73} 3195 (1994).
%
\bibitem{Elgaroy03} O. Elgaroy, S. Hannestad, and T. Haugboelle, JCAP, 09, 008 (2003).
%
\bibitem{Romano08} A. E. Romano and M. Sasaki, Phys. Rev. D {\bf 78}, 103522 (2008).
%
\bibitem{Barnaby09} N. Barnaby, Z. Huang, L. Kofman, and D. Pogosyan, Phys. Rev. D {\bf 80}, 043501 (2009).
%
\bibitem{Fedderke15} M. A. Fedderke, E. W. Kolb, M. Wyman, Phys. Rev.,  D {\bf 91}, 063505 (2015).
%
\bibitem{Starobinsky02} A. A. Starobinsky and I. I. Tkachev, J. Exp. Th. Phys.  Lett., {\bf 76}, 235 (2002).
%
\bibitem{Camb}
A. Lewis, A. Challinor,  and A. Lasenby, Astrophys. J., {\bf 538}, 473 (2000).
%
\bibitem{Christensen}
N. Christensen and R. Meyer, L. Knox, and B. Luey,  
Class. and Quant.  Grav., {\bf 18}, 2677 (2001).
%
\bibitem{Lewis}
A. Lewis and S. Bridle, Phys. Rev. D {\bf 66}, 103511 (2002).
%
\bibitem{birrellanddavies} N. D. Birrell and P. C. W. Davies, Quantum
Fields in Curved Space, (Cambridge Univ. Press, Cambridge, 1982).
%
\bibitem{Kofman:1997yn}
L.~Kofman, A.~Linde and A.~A.~Starobinsky,
Phys.\ Rev.\ D {\bf 56}, 3258 (1997).
%
\bibitem{Chung:1998bt}
D.~J.~H.~Chung,
Phys.\ Rev.\ D {\bf 67}, 083514 (2003).
%
\bibitem{Kolb84} E. W. Kolb and R. Slansky, Phys. Lett 135B, 378 (1984).
%
\bibitem{Lewis03} {\it Our Superstring Universe: Strings, Branes, Extra Dimensions and Superstring-M Theory }, by  L.E. Lewis, Jr.
  (iUniverse, Inc. NE, USA;  2003)
    
\bibitem{efstathiou} G. P. Efstathiou, in {\it Physics of the Early
Universe}, (SUSSP Publications, Edinburgh, 1990), eds. A. T. Davies,
A. Heavens, and J. Peacock.
%
\bibitem{Peacock}
J. A. Peacock and S. J. Dodds, Mon. Not. R. Astron. Soc., 
{\bf 280} L19 (1996).
%
\bibitem{wiggle} E. A. Kazin, et al. (Wiggle-Z Dark Energy Survey), MNRAS {\bf 441}, 3524-3542 (2014).
%
\bibitem{Silverstein08}E. Silverstein, and A.  Westphal,  Phys.Rev., {\bf D78}, 106003 (2008).
%
\bibitem{McAllister10} L. McAllister, E. Silverstein and  A.  Westphal, Phys.Rev., {\bf D82}, 046003 (2010).
%
\end{thebibliography}
\end{document}